# EXPERIMENTS ON PROPAGATING AND BRANCHING POSITIVE STREAMERS IN AIR

T M P Briels[1], E M van Veldhuizen[1], U Ebert[1,2]

[1]Department of Applied Physics, Eindhoven University of Technology,
P.O. Box 513, 5600 MB Eindhoven, The Netherlands, t.m.p.briels@tue.nl
[2]Institute for Mathematics and Computer Science (CWI),
P.O. Box 94079, 1090 GB Amsterdam, The Netherlands, ute.ebert@cwi.nl

## ABSTRACT
The evolution of streamers is known to depend on gas composition and electrode geometry, on polarity and size of the voltage and also on the electric circuit that produces the high voltage pulse. To characterize the phenomena better and to compare with theory, a new larger experimental setup with vacuum enclosure has been built. We here present first results in this setup on positive streamers in air at fixed voltage and varying electrode distance. While next to the emitting anode tip, a similar number of streamers seems to emerge due to multiple branching, more streamers seem to survive over a fixed distance, when the gap is shorter. When lowering the voltage, streamers branch less at all distances from the anode tip or do not branch at all beyond a certain distance.

## INTRODUCTION
When suddenly a high electric field is applied to a non-conducting medium, electric breakdown might concentrate in narrow, rapidly growing ionised channels, so called streamers. These streamers are observed in gases as well as in liquids and in non-conducting solids. They evolve within tens to hundreds of nanoseconds, and often they branch. Streamers form the early stages of sparks of all sizes, from spark plugs up to large-scale discharges in the atmosphere.

Even though streamers are frequent in nature and used in a wide range of applications, their rapid propagation as well as their complex dynamic and nonlinear nature has long hindered a full understanding. The present paper makes part of a new initiative that coordinates experiment and theory.

Some recent results: Recent numerical work deals with anode directed (negative) streamers in air [1] and or in simple gases like nitrogen and argon [2,3]. Cathode directed (positive) streamers are modelled in air [1,4] for a 10 mm gap and nitrogen/oxygen mixtures [5] for a 130 mm gap. Experiments are performed on positive streamers in an 11 mm [6] and 25 mm gap in air and argon [7-9] and in pure nitrogen and oxygen [9]. In a large gap of 130 mm, negative and positive streamers in nitrogen/oxygen mixtures have been investigated experimentally for different voltages [10]. The properties of an individual streamer propagating simultaneously with others in air are investigated experimentally in [6].

An important experimental fact is that streamers can branch many times. The first quantitative prediction of streamer branching can be found in [2] for negative streamers in strong fields. The numerical evidence of streamer branching [2,3] is presently elaborated with improved computations [11] as well as with analytical theory [2,3,12,13].

In this paper, the influence of gap distance and applied peak voltage on the evolution and branching of streamer channels is discussed. All experiments are performed in the same setup; only the electrode distance and peak voltage is adjusted.

## EXPERIMENTAL OVERVIEW
### Setup
In Figure 1, a scheme of the electric circuit is shown. The high voltage power supply (up to 30 kV) charges the capacitor $C$ (1.3 nF) via resistor $R_1$ (25 M$\Omega$). When the high voltage semiconductor switch (Behlke HTS-301) is closed, the circuit is short-circuited and a positive pulse is applied to the anode-tip. This tip is made of tungsten with a small amount of thorium and has a very sharp tip to create an electric field as high as possible. At this spot, a corona discharge will take place in the direction of the planar cathode, which is connected to earth. The anode and cathode are located inside a grounded vacuum vessel.

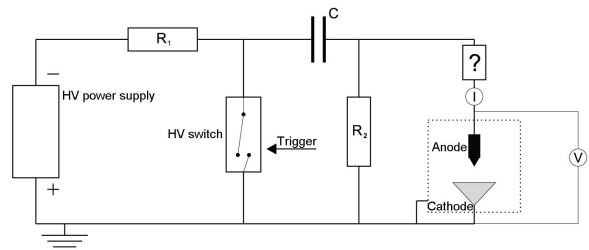

*Figure 1: Electric circuit for pulses up to 30 kV.*

The resistor $R_2$ (25 M$\Omega$) reduces the voltage at the anode to zero after a discharge. At the spot of the question mark, different components (e.g. a wire, resistor or inductance) can be placed. In the current experiments a wire is used. The electrode configuration is point-plane and the distance between the electrodes is adjustable.

The current and high-voltage probes (Pearson 2887 and Tektronix P6105A, respectively) are attached to an oscilloscope (Tektronix TDS 380), to be able to measure *I-V*-characteristics on nanosecond time scale. Pictures of the streamers are made with a fast ICCD-camera (Andor DH534-18), which has a minimal optical gate of 0.8 ns.

**Results and discussion**

All measurements are done on positive streamers in air at atmospheric pressure. They show that the branching pattern depends on the distance between the electrodes and on the applied peak voltage, see Figures 2-4. The pictures are time integrated (20 μs). In all pictures, the anode tip is on top of the picture (where the streamers start) and the cathode plate is at the bottom of the picture, as indicated in Figure 2. A number of branches may not show up on the picture since the camera's focus is in the plane of the anode tip. The camera's sensitivity is optimised to show streamer branches, hence the streamers at the anode tip can be overexposed.

The number of branches observed for positive streamers in air and their propagation length, is in general quite reproducible within each used setting. The distance that the streamers propagate depends on the applied voltage. When comparing the pictures at 17.5 kV (Figure 2), 12.5 kV (Figure 3) and 7.5 kV (Figure 4), it is clear that in general at higher voltages more streamers bridge the gap and more branches are observed. This was already shown in [7]. The lowest voltage at which the streamers reach the cathode is approximately 6.5 kV at a 10 mm gap.

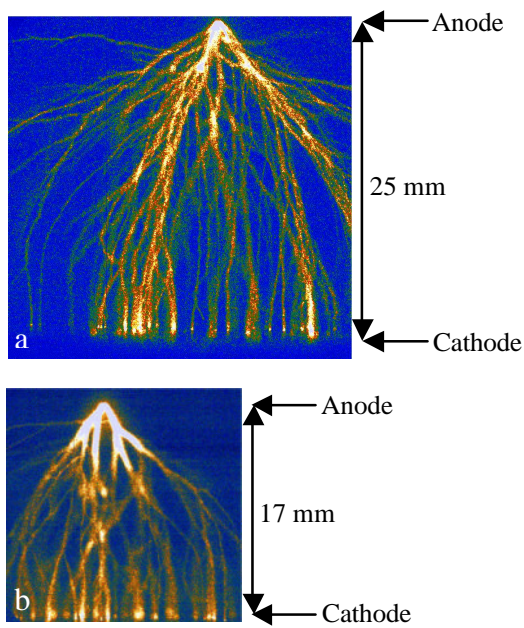

*Figure 2: Positive streamers in air at 17.5 kV, in a point-plane geometry with a gap of 25 mm (a) and 17 mm (b). (a) shows ~70 branches and (b) ~40.*

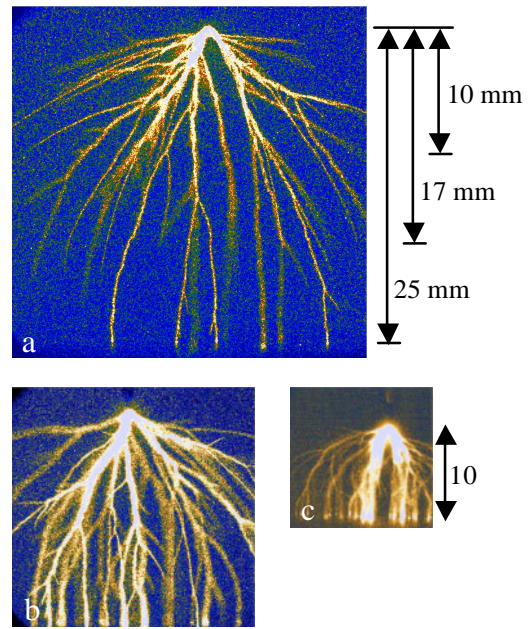

*Figure 3: Same setup as in Fig. 2, now at 12.5 kV. a) 25 mm, ~45 branches; b) 17 mm, ~40 branches; c) 10 mm, ~30 branches.*

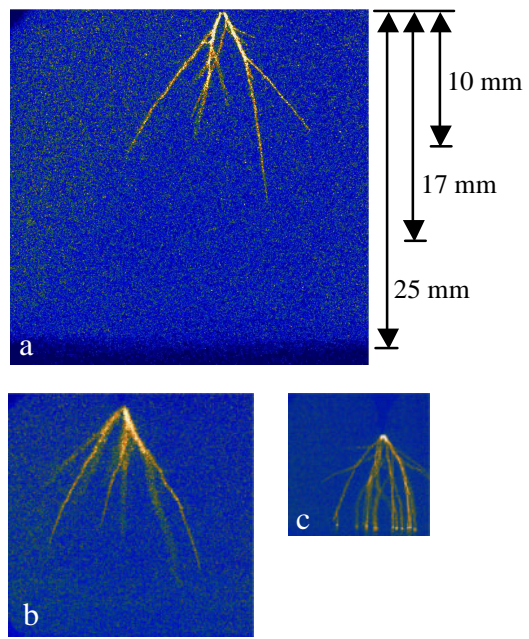

*Figure 4: Same setup as in Fig. 2, now at 7.5 kV. a) 25 mm, 8 branches; b) 17 mm, 8 branches; c) 10 mm, ~20 branches.*

Comparing the subfigures a and b of Figures 2 and 4, the branching frequency looks quite similar close to the emitting anode tip. This can be understood from an electrostatic consideration: The local electric fields close to the tip are determined primarily by strongly curved parts of the anode tip and the applied voltage and not so much by the distance to the planar electrode, if it is far enough away. Of course, further away from the tip, the background field does depend on the gap length, but arriving there, the streamer generates its own electric field enhancement at its tip. For the short gap of Figures 3c and 4c, more branching is observed.

Voltages and gap lengths are chosen such that different figures have the same average electric field: Figures 2a, 3b and 4c have an average field of ~7.5 kV/cm, Figures 3a and 4b of ~5 kV/cm. Obviously the average electric field strength is not a good parameter, since the pictures are quite different.

The pictures also show that at low voltages, the streamers branch mostly in the first 5 mm from the anode (Figure 4). At large voltages, the streamers branch in the first 10 mm (Figure 2).

The *I-V*-characteristic can also be analysed. The current, which flows in the loop of the capacitor, the switch and the discharge gap, consists of two parts (Figure 6). The first part is a displacement current, $I_{displ}$, which brings the needle to a high potential, given by

$$I_{displ} = C_{geom}\frac{dV}{dt} \qquad (1)$$

where $C_{geom}$ is the total geometrical capacitance of the electrode configuration. The second part is the real discharge current, $I_{disch}$. $C_{geom}$ is calculated by fitting the calculated d$V$/d$t$, the dotted line in Figure 6, to the first current peak. This gives a value of ~13.8 pF. From the second part, called discharge in Figure 6, the energy of the streamers is calculated. Here fore, the average current difference between the measured current (solid line) and displacement current (dotted line) is multiplied by the voltage and the duration of the discharge. This gives a value of 1.7 mJ. This is comparable to [7] where an energy of 2 mJ for ~100 branches is found for a 25 kV pulse in a 25 mm gap. From these results it is estimated that there are ~50 branches/mJ. Therefore, for the discharge in Figure 2a an energy of ~2mJ is estimated. For the Figures 2b and 3, the energy is ~1 mJ, in Figure 4a,b it is ~0.1 mJ and in 4c it is ~0.2 mJ.

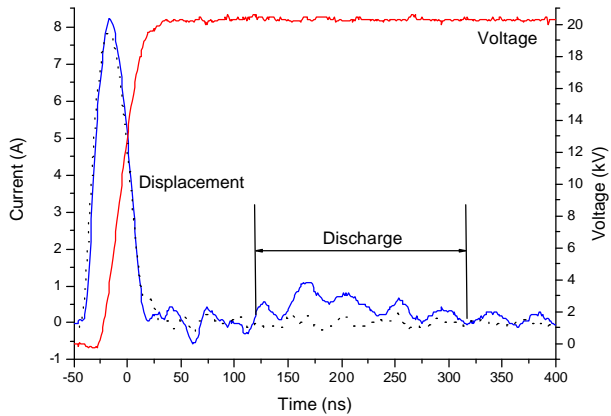

*Figure 6: I-V characteristic for a discharge in a 17 mm point-plane gap in air at 20 kV. One solid line is the voltage, the other one is the measured current. The dotted line represents the displacement current.*

## THEORETICAL OVERVIEW

In parallel to these experimental investigations, in Amsterdam, analytical [2,3,12,13] and numerical work [2,3,11] is done on streamer propagation. The numerical fluid model is based on the least number of physical processes that are necessary for negative streamer formation. It assumes that free electrons and positive ions are generated by impact of accelerated electrons on neutral molecules. The ion mobility is neglected. It is appropriate for negative streamers in simple non-attaching gases like argon or nitrogen. In dimensionless units, the model has the form:

$$\partial_t \boldsymbol{s} - \nabla \cdot (\boldsymbol{s}E + D\nabla \boldsymbol{s}) = \boldsymbol{s}f(|E|) \qquad (2)$$

$$\partial_t \boldsymbol{r} = \boldsymbol{s}f(|E|) \qquad (3)$$

$$\boldsymbol{r} - \boldsymbol{s} = \nabla \cdot E \qquad (4)$$

$$E = -\nabla \boldsymbol{F} \qquad (5)$$

$$f(|E|) = |E|\boldsymbol{a}(|E|) \qquad (6)$$

where $\boldsymbol{s}$ is the density of electrons, $\boldsymbol{r}$ the density of positive ions, $E$ the electric field, $D$ the diffusion constant, $\boldsymbol{F}$ the electric potential and $\boldsymbol{a}$ the effective cross section of impact ionisation. Photo-ionisation and electron attachment will be included in future investigations.

Numerical solutions in a strong homogeneous electric field show that instabilities arise spontaneously (Figure 7). This can be considered as the first step of branching. It is observed here in a 3-D setting with assumed cylindrical symmetry. This symmetry makes the later solutions after branching unphysical, but it proves the presence of the instability also in a fully 3-D setting. The instability mechanism is an interfacial Laplacian instability of the head of the highly conducting channel [2,3,13].

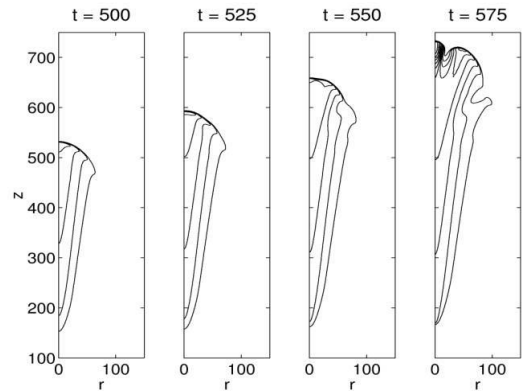

*Figure 7: Evolution of an anode directed streamer in a strong homogeneous background field of 100 kV/cm [3]. In dimensionless coordinates, the planar cathode is located at z = 0 and the planar anode at z = 2000, corresponding to 4.6 mm. The lines denote levels of equal electron density $\boldsymbol{s}$ with increments of 0.15.*

# CONCLUSIONS

Pulsed corona streamers in a short gap in air are, considering streamer number, length and diameter, quite reproducible. Their propagation and branching depends on the applied peak voltage and the distance between the electrodes. At higher electric fields, more streamer branches arise. This is shown when the voltage is increased and the electrode distance is fixed. It also occurs when the voltage is fixed and the distance, travelled by the streamers, in different gaps is compared. However, the streamer patterns at similar average electric field strengths, but different gap lengths, have no resemblance. At least close to a pointed electrode, the local field strength is the physically important parameter.

Even though the streamers branch throughout the whole gap, they seem to branch mostly close to the pointed anode in the high field area. At low applied voltages this occurs within the first 5 mm, while at higher voltages it occurs within the first 10 mm.

The numerical model also shows instabilities, which can be considered as the first step of branching.

# FUTURE

The aim of the investigation is a quantitative comparison between theory and experiment of streamer width, velocity, intensity and branching statistics. At the moment, experiments and theory differ by sign of discharge, gas composition, electrode geometry, voltage pulse shape and field strength. With the new setup a wide range of parameters will be explored for positive and negative streamers in different gases at different pressures, different anode-cathode distances and configurations and different voltages, pulse forms and discharge triggering.

The numerical model will be extended with photo-ionisation and electron attachment.

# REFERENCES

[1] G.E. Georgiou, R. Morrow, A.C. Metaxas, "Two-dimensional simulation of streamers using the FE-FCT algorithm", J. Phys. D.: Appl. Phys. **33**, L27-L32 (2000).
[2] M. Arrayás, U. Ebert, W. Hundsdorfer, "Spontaneous branching of anode-directed streamers between planar electrodes", Phys. Rev. Lett. **88**, 174502 (2002).
[3] A. Rocco, U. Ebert, W. Hundsdorfer, "Branching of negative streamers in free flight", Phys. Rev. E **66**, 035102(R) (2002).
[4] A.A. Kulikovsky, "Positive streamer in a weak field in air: A moving avalanche-to-streamer transition", Phys. Rev. E **57**, 7066-7074 (1998).
[5] S.V. Pancheshnyi, A.Yu. Starikovskii, "Two-dimensional numerical modelling of the cathode-directed streamer development in a long gap at high voltage", J. Phys. D.: Appl. Phys. **36**, 2683-2691 (2003).
[6] P. Tardiveau, E. Marode, A. Agneray, "Tracking an individual streamer branch among others in a pulsed induced discharge", J. Phys. D.: Appl. Phys. **35**, 2823-2829 (2002).
[7] E.M. van Veldhuizen, W.R. Rutgers, "Pulsed positive corona streamer propagation and branching", J. Phys. D.: Appl. Phys. **35**, 2169-2179 (2002).
[8] E.M. van Veldhuizen, P.C.M. Kemps, W.R. Rutgers, "Streamer branching in a short gap: the influence of the power supply", IEEE Trans. Pl. Sc. **30**, 162-163 (2002).
[9] E.M. van Veldhuizen, W.R. Rutgers, "Inception behaviour of pulsed positive corona in several gases", J. Phys. D.: Appl. Phys **36**, 2692-2696 (2003).
[10] W.J. Yi, P.F. Williams, "Experimental study of streamers in pure $N_2$ and $N_2/O_2$ mixtures and a ≈13 cm gap", J. Phys. D.: Appl. Phys **35**, 205-218 (2002).
[11] C. Montijn, J. Wackers, W. Hundsdorfer, U. Ebert, in preparation.
[12] M. Arrayas, U. Ebert, "Stability of negative ionization fronts: Regularization by electric screening?", Phys. Rev. E **69**, 036214 (2004).
[13] B. Meulenbroek, A. Rocco, U. Ebert, "Streamer branching rationalized by conformal mapping techniques", Phys. Rev. E **69**, 04???? (2004) [to be published in April].
[14] E.M. van Veldhuizen (editor), "Electrical discharges for environmental purposes: fundamentals and applications", Nova Science Publishers Inc., New York (2000).